\documentclass[aps,graphicx,6pt,showkeys,showpacs,onecolumn]{revtex4}
\usepackage{amsmath}
\usepackage{amscd}
\usepackage{graphicx}
\usepackage{subfigure}
\usepackage{appendix}

\begin{document}

\title[Short Title]{Shortcuts to adiabatic passage for fast generation of Greenberger-Horne-Zeilinger states by transitionless quantum driving}

\author{Ye-Hong Chen$^{1}$}
\author{Yan Xia$^{1,}$\footnote{E-mail: xia-208@163.com}}
\author{Jie Song$^{2,}$\footnote{E-mail: jsong@hit.edu.cn}}
\author{Qing-Qin Chen$^{3}$}

\affiliation{$^{1}$Department of Physics, Fuzhou University, Fuzhou
350002, China\\$^{2}$Department of Physics, Harbin Institute of
Technology, Harbin 150001, China\\$^{3}$Zhicheng College, Fuzhou University, Fuzhou
350002, China}


\begin{abstract}
Berry's approach on ``transitionless quantum driving'' shows how to
set a Hamiltonian which drives the dynamics of a system along
instantaneous eigenstates of a reference Hamiltonian to reproduce
the same final result of an adiabatic process in a shorter time. In
this paper, motivated by transitionless quantum driving, we
construct shortcuts to adiabatic passage in a three-atom system to
create the Greenberger-Horne-Zeilinger states with the help of
quantum Zeno dynamics and of non-resonant lasers. The influence of
various decoherence processes is discussed by numerical simulation
and the result proves that the scheme is fast and robust against
decoherence and operational imperfection.
\end{abstract}


\maketitle

\small

\section{Introduction}
``Shortcuts to adiabatic passage (STAP)''
\cite{XCILARDGOJGMPra10,ETSISMGMMACDGOARXCJGMAmop13}
which are a set of techniques to speeding up a slow quantum
adiabatic process usually through a non-adiabatic route,
have attracted a great deal of attention in recent years.
They can overcome the harmful effects caused by decoherence, noise or losses
during a long operation time. Quantum science also greatly desires fast and robust theoretical methods
since high repetition rates contribute to the achievement of better signal-to-noise ratios and better accuracy.
Therefore, in the last several years, STAP
have been applied in a wide range of systems in theory and experiment
\cite{AdCPrl13,SMKNPrspca10Pra11,XCARSSADCDDOJGMPrl10,YHCYXQQCJSPra14,MLYXLTSJSLp14,YHCYXQQCJSLpl14,MLYXLTSJSNBAPra14,ARXCDAJGMNjp12,YHYXQQCJSarxiv14,
JFSXLSPVGLPra10,JFSXLSPCPVGLEpl11,AWFZTRSTDOKTMHKSFSKUPPrl12,SYTXCOl12,JGMXCARDGOJpb09,XCJGMPra10,JFSPCGLPVNjp11,
ETSIXCARDGOJGMPra11,XCETDSJSLJGMPra11,ETXCMMSSARJGMNjp12,YLLAWZDWPra11,AdCPra11}.
Various reliable, fast and robust methods and schemes have been
proposed to implement quantum information processing (QIP), such as
fast population transfer
\cite{YHCYXQQCJSPra14,MLYXLTSJSNBAPra14,MLYXLTSJSLp14}, fast
entanglement generation \cite{YHCYXQQCJSLpl14,MLYXLTSJSNBAPra14},
fast implementation of quantum phase gates \cite{YHYXQQCJSarxiv14}.

To construct shortcuts to speed up adiabatic processes effectively,
two methods which are in fact strongly related, and even potentially
equivalent to each other \cite{XCETJGMPra10}: are invariant-based
inverse engineering based on Lewis-Riesenfeld invariant
\cite{JGMXCARDGOJpb09,HRLWBRJmp69} and  Berry's approach named
``transitionless quantum driving'' (TQD)
\cite{MVBJpa09,MGBMVNMPHEADCRFVGRMONatPhys12,MDSARJpca03,MDSARJcp08}.
Whereas, each of the two methods also has its own characteristics,
for example, using Lewis-Riesenfeld invariants to construct
shortcuts usually does not have to break down the form of the
original Hamiltonian $H_{0}(t)$, so that the possibility of
designing a Hamiltonian $H(t)$ very difficult or impossible to
implement in practice is avoided \cite{XCJGMPra10,YHCYXQQCJSPra14}.
However, the invariants always have fixed forms which lead to that
shortcut methods based on Lewis-Riesenfeld invariants might be
limited or even hopeless in some cases to construct shortcuts to
implement QIP rapidly \cite{YHCYXQQCJSPra14}. For example, in the
paper \cite{YHCYXQQCJSLpl14} proposed by Chen \emph{et al.}, they
had no choice but to make one of the atoms to be a control qubit or
use auxiliary levels for the atoms to generate entangled states.

There is still plenty to do to make wide applications of STAP for
fast QIP in some experimental systems, for example, the cavity
quantum electronic dynamics (QED) systems. It is worth noting that,
TQD provides a very
effective method to construct the ``counter-diabatic
driving'' (CDD) Hamiltonian $H(t)$ which accurately
drives the instantaneous eigenstatees of $H_{0}(t)$.
Nevertheless, it is almost always found that the designed CDD Hamiltonian is hard to
be directly implemented in practice, especially in multiparticle systems. 
Examples of ways to overcome this problem may be found in
Ref.
\cite{SIXCETJGMARPrl12,SMGETXCJGMPra14,ETSMGJGMPra14,TOKMMjp14}.
Also, in a large detuning limit, Lu \emph{et al.}
\cite{MLYXLTSJSNBAPra14}  have  found a simplified effective
Hamiltonian equivalent to $H(t)$. This idea inspires us that finding
an alternative physically feasible (APF) Hamiltonian which is
effectively equivalent to $H(t)$. 
However, the approximation in ref.
\cite{MLYXLTSJSNBAPra14} is too complex to be generalized to
$N$-qubit entanglement cases. It is known to all that, entanglement
of more qubits shows more nonclassical effects and is more useful
for quantum applications. For example, one of the two kinds of
three-qubit entangled states named the Greenberger-Horne-Zeilinger
(GHZ) states  provide a possibility for testing quantum mechanics
against local hidden theory without using Bell's inequality
\cite{DMGMHASAZAjp90,SBZPrl01}. Therefore, great interest has arisen
regarding the significant role of the GHZ states in the foundations
of quantum mechanics measurement theory and quantum communication.
In view of that we wonder if it is possible to use TQD to construct
shortcuts for one-step generation of multi-qubit entanglement, i.e.,
the three-atom GHZ states, without abandoning any of the atoms or
using auxiliary levels.

In this scenario, motivated by refs.
\cite{YHCYXQQCJSPra14,YHCYXQQCJSLpl14,MLYXLTSJSNBAPra14,MLYXLTSJSLp14},
we use TQD to construct STAP to generate the three-atom GHZ states
effectively and rapidly in one step. It would be a promising idea of
applying STAP to realize multi-qubit entanglement generation in
cavity QED systems. Different from ref. \cite{MLYXLTSJSNBAPra14}, we
use the quantum Zeno dynamics \cite{PKHWTHAZMAKPrl95,PFSPPrl02} to
simplify the system first and then under the large detuning
conditon, we obtain the effective Hamiltonian which is
equivalent to the corresponding CDD Hamiltonian to speed up the
evolution process.
Therefore, the adiabatic process for a multi-qubit system is speeded
up, and the STAP is easy to be achieved in experiment. Comparing
with ref. \cite{YHCYXQQCJSLpl14}, we use TQD in this paper so that
the laser pulses are not strongly limited and we do not need to use
auxiliary levels or multi-step operations to generate the three-atom
GHZ states. Moreover, we find that any quantum system whose
Hamiltonian is possible to be simplified into the form in eq.
(\ref{eq1-4}), the corresponding APF Hamiltonian can be built and
the STAP can be constructed with the same approach presented in this
paper. The above advantages mean the present scheme is much more
useful in dealing with the fast and noise-resistant generation of
multi-qubit entanglement or even other QIP.

The paper is structured as follows. We first give a brief
description about the quantum Zeno dynamics and the approach of TQD
proposed by Berry in section II. In section III, we describe a
theoretical model 
for three $\Lambda$-type atoms
which are trapped in a bimodal-mode cavity. In section IV, we show
how to construct STAP for the system in section III. In section V,
we use the constructed shortcut to generate a three-atom GHZ state
and give the numerical simulation and experimental discussion about
the validity of the scheme. Finally, in section VI, the conclusion
is given.

\begin{figure}
 \scalebox{0.45}{\includegraphics {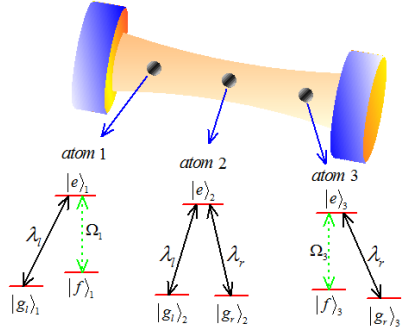}}
 \caption{The cavity-atom combined system and the atomic level configuration for the original Hamiltonian.}
 \label{model}
\end{figure}

\begin{figure}
 \scalebox{0.45}{\includegraphics {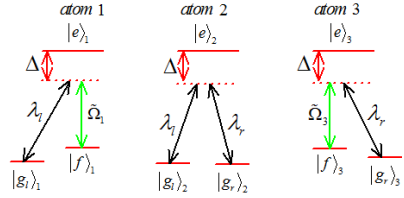}}
 \caption{The atomic level configuration for the APF Hamiltonian.}
 \label{modelGHZ2}
\end{figure}

\section{Basic theories}
\subsection{Transitionless quantum driving}
Consider an arbitrary time-dependent Hamiltonian $H_{0}(t)$, with instantaneous eigenstates and
energies given by
\begin{eqnarray}\label{eq0-1}
  H_{0}(t)|\varphi_{n}(t)\rangle=\zeta_{n}(t)|\varphi_{n}(t)\rangle.
\end{eqnarray}
When this system satisfies the adiabatic condition, $H_{0}(t)$ will drive the system into
\begin{eqnarray}\label{eq0-2}
  |\psi(t)\rangle=e^{i\vartheta_{n}(t)}|\varphi_{n}(t)\rangle,
\end{eqnarray}
where
\begin{eqnarray}\label{eq0-3}
  \vartheta_{n}(t)=-\frac{1}{\hbar}\int_{0}^{t}{dt' \zeta_{n}(t')}
                  +i\int_{0}^{t}{dt'\langle\varphi_{n}(t')|\partial_{t'}\varphi_{n}(t')\rangle}.
\end{eqnarray}
To find the
Hamiltonian $H(t)$ that drives the eigenstates
$\{|\varphi_{n}(t)\rangle\}$, we define a unitary operator
\begin{eqnarray}\label{eq0-4}
  U=\sum_{n}e^{i\vartheta_{n}(t)}|\varphi_{n}(t)\rangle\langle\varphi_{n}(0)|,
\end{eqnarray}
which obeys
\begin{eqnarray}\label{eq0-5}
  i\hbar\partial_{t}U=H(t)U\Rightarrow H(t)=i\hbar(\partial_{t}U)U^{\dag}.
\end{eqnarray}
Then the Hamiltonian $H(t)$ is obtained
\begin{eqnarray}\label{eq0-6}
  H(t)&=&H_{0}(t)+H_{1}(t),             \cr\cr
  H_{1}(t)&=&i\hbar\sum_{n}{(|\partial_{t}\varphi_{n}\rangle\langle\varphi_{n}|
           -\langle\varphi_{n}|\partial_{t}\varphi_{n}\rangle|\varphi_{n}\rangle\langle\varphi_{n}|)}.
\end{eqnarray}
The simplest choice is $\zeta_{n}=0$, for which the bare states $|\varphi_{n}(t)\rangle$, with no phase factors, are driven by \cite{MVBJpa09}
\begin{eqnarray}\label{eq0-7}
  H(t)=i\hbar\sum_{n}{|\partial_{t}\varphi_{n}\rangle\langle\varphi_{n}|},
\end{eqnarray}
reflecting
\begin{eqnarray}\label{eq0-8}
  i\hbar\partial_{t}|\varphi_{n}\rangle=i\hbar\sum_{m}|\partial_{t}\varphi_{m}\rangle\langle\varphi_{m}|\varphi_{n}\rangle.
\end{eqnarray}
\subsection{Quantum Zeno dynamics}
The quantum Zeno dynamics was named by Facchi and Pascazio in 2002
\cite{PFSPPrl02}. It is derived from the quantum Zeno effect which
describes a phenomenon that the system can actually evolve away from
its initial state while it still remains in the so-called Zeno
subspace determined by the measurement when frequently projected
onto a multidimensional subspace. According to von Neumann's
projection postulate, the quantum Zeno dynamics can be achieved via
continuous coupling between the system and an external system
instead of discontinuous measurements \cite{PFSPPrl02}. In general,
we assume that a dynamical evolution process is governed by the
Hamiltonian
\begin{eqnarray}\label{eq0b-1}
  H_{Z}=H_{obs}+KH_{meas},
\end{eqnarray}
where $H_{obs}$ is the Hamiltonian of the quantum system investigated, $K$ is a coupling constant, and $H_{meas}$
is viewed as an additional interaction Hamiltonian performing the measurement.
In the ``infinitely strong measurement'' limit $K\rightarrow\infty$ \cite{PKHWTHAZMAKPrl95,PFSPPrl02},
The Hamiltonian for the whole system is nearly equivalent to
\begin{eqnarray}\label{eq0b-2}
  H_{Zeno}=\sum_{n}(P_{n}H_{obs}P_{n}+\varepsilon_{n}P_{n}),
\end{eqnarray}
whit $P_{n}$ being the $n$th orthogonal projection onto the invariant Zeno subspace $\forall_{Pn}$
and the eigenspace of $KH_{meas}$ belonging to the eigenvalue $\varepsilon_{n}$, i.e., $KH_{meas}P_{n}=\varepsilon_{n}P_{n}$.

\section{Model}
We consider three $\Lambda$-type atoms are trapped in a bimodal-mode
cavity as shown in Fig. \ref{model}. Atoms $1$, $2$, and $3$ have
three sets of ground states \{$|f\rangle_{1}$,
$|g_{l}\rangle_{1}$\}, \{$|g_{l}\rangle_{2}$, $|g_{r}\rangle_{2}$\},
and \{$|f\rangle_{3}$, $|g_{r}\rangle_{3}$\}, respectively, and each
of them has an excited state $|e\rangle$. The atomic transition
$|f\rangle\leftrightarrow|e\rangle$ is driven resonantly through
classical laser field with time-dependent Rabi frequency
$\Omega(t)$, transition $|g_{l}\rangle\leftrightarrow|e\rangle$ is
coupled resonantly to the left-circularly polarized mode of the
cavity with coupling $\lambda_{l}$, and transition
$|g_{r}\rangle\leftrightarrow|e\rangle$ is coupled resonantly to the
right-circularly polarized mode of the cavity with coupling
$\lambda_{r}$. Under the rotating-wave approximation (RWA), the
interaction Hamiltonian for this system reads ($\hbar=1$):
\begin{eqnarray}\label{eq1-1}
  H_{I}&=&H_{al}+H_{ac},                                                  \cr\cr
  H_{al}&=&\Omega_{1}(t)|e\rangle_{1}\langle f|+e^{i\beta}\Omega_{3}(t)|e\rangle_{3}\langle f|+H.c.,       \cr\cr
  H_{ac}&=&\sum_{k=1,2}{\lambda_{l}a_{l}|e\rangle_{m}\langle g_{l}|}+\sum_{j=2,3}{\lambda_{r}a_{r}|e\rangle_{n}\langle g_{r}|}+H.c.,
\end{eqnarray}
where $a_{l}$ and $a_{r}$ are the left- and right-circularly annihilation operators of the cavity modes,
and $\beta$ means the two Rabi frequencies are $\beta$-dephased from each other.
If we assume the initial state is $|f,g_{l},g_{r}\rangle_{1,2,3}|0,0\rangle_{c}$, the system will evolve within a single-excitation subspace with basis
states
\begin{eqnarray}\label{eq1-2}
  |\psi_{1}\rangle&=&|f,g_{l},g_{r}\rangle_{1,2,3}|0,0\rangle_{c},      \cr
  |\psi_{2}\rangle&=&|e,g_{l},g_{r}\rangle_{1,2,3}|0,0\rangle_{c},      \cr
  |\psi_{3}\rangle&=&|g_{l},g_{l},g_{r}\rangle_{1,2,3}|1,0\rangle_{c},  \cr
  |\psi_{4}\rangle&=&|g_{l},e,g_{r}\rangle_{1,2,3}|0,0\rangle_{c},      \cr
  |\psi_{5}\rangle&=&|g_{l},g_{r},g_{r}\rangle_{1,2,3}|0,1\rangle_{c},  \cr
  |\psi_{6}\rangle&=&|g_{l},g_{r},e\rangle_{1,2,3}|0,0\rangle_{c},      \cr
  |\psi_{7}\rangle&=&|g_{l},g_{r},f\rangle_{1,2,3}|0,0\rangle_{c}.
\end{eqnarray}
In light of quantum Zeno dynamics, we rewrite the Hamiltonian $H_{I}$ in eq. (\ref{eq1-1}) as $H_{re}$ through
the relation $H_{re}=\sum_{n}P_{n}H_{al}P_{n}+\varepsilon_{n}P_{n}$ ($H_{al}\rightarrow H_{obs}$ and $H_{ac}\rightarrow KH_{meas}$), where
\begin{eqnarray}\label{eq1-2a}
  P_{n}&=&\sum_{m}{|m\rangle\langle m|},         \cr
  |m\rangle&\in&\{|\psi_{1}\rangle,|\psi_{7}\rangle,|\phi_{0}\rangle, |\phi_{1}\rangle, |\phi_{2}\rangle, |\phi_{3}\rangle, |\phi_{4}\rangle\}.
\end{eqnarray}
Here $|\phi_{0}\rangle$, $|\phi_{1}\rangle$, $|\phi_{2}\rangle$, $|\phi_{3}\rangle$, and $|\phi_{4}\rangle$
are the eigenvectors of $H_{ac}$ corresponding eigenvalues $\varepsilon_{0}=0$, $\varepsilon_{1}=\lambda$, $\varepsilon_{2}=-\lambda$,
$\varepsilon_{3}=\sqrt{3}\lambda$, and $\varepsilon_{4}=-\sqrt{3}\lambda$, respectively.
And we obtain (we set $\lambda_{l}=\lambda_{r}=\lambda$)
\begin{eqnarray}\label{eq1-3}
  H_{re}&=&\sum_{k=0}^{4}{\varepsilon_{k}|\phi_{k}\rangle\langle\phi_{k}|}+H_{al}^{re},                                                          \cr\cr
  H_{al}^{re}&=&\frac{1}{\sqrt{3}}{[|\phi_{0}\rangle+\frac{1}{2}(|\phi_{3}\rangle+|\phi_{4}\rangle)]
           (\Omega_{1}\langle\psi_{1}|+e^{i\beta}\Omega_{3}\langle\psi_{7}|)}                                                                \cr\cr
           &&+\frac{1}{2}(|\phi_{1}\rangle+|\phi_{2}\rangle)(-\Omega_{1}\langle\psi_{1}|+e^{i\beta}\Omega_{3}\langle\psi_{7}|)               \cr\cr
           &&+H.c..
\end{eqnarray}
Through performing the unitary transformation
$U_{Z}=e^{-i\sum{\varepsilon_{k}|\phi_{k}\rangle\langle\phi_{k}|}t}$
and neglecting the terms with high oscillating frequency by setting
the condition $\Omega_{1}/\sqrt{3},\Omega_{3}/\sqrt{3}\ll \lambda$
(the Zeno condition), we obtain an effective Hamiltonian
\begin{eqnarray}\label{eq1-4}
  H_{eff}=\frac{1}{\sqrt{3}}|\phi_{0}\rangle(\Omega_{1}(t)\langle\psi_{1}|+e^{i\beta}\Omega_{3}(t)\langle\psi_{7}|)+H.c.,
\end{eqnarray}
which can be seen as a simple three-level system with an excited state $|\phi_{0}\rangle$ and two ground states $|\psi_{1}\rangle$ and $|\psi_{7}\rangle$.
For this effective Hamiltonian, its eigenstates are easily obtained
\begin{eqnarray}\label{eq1-5}
  |n_{0}(t)\rangle&=&
    \left(
     \begin{array}{c}
         \cos{\theta(t)}      \\
        0                     \\
        -e^{i\beta}\sin{\theta(t)}      \\
     \end{array}
    \right),   \cr\cr \cr
  |n_{\pm}(t)\rangle&=&
    \frac{1}{\sqrt{2}}\left(
     \begin{array}{c}
       \sin{\theta(t)}    \\
       \pm1               \\
       e^{i\beta}\cos{\theta(t)}   \\
     \end{array}
    \right),
\end{eqnarray}
corresponding eigenvalues $\eta_{0}=0$,
$\eta_{\pm}=\pm\Omega/\sqrt{3}$, respectively, where
$\tan{\theta}=\Omega_{1}/\Omega_{3}$ and
$\Omega=\sqrt{(\Omega_{1}^{2}+\Omega_{3}^{2})}$. When the adiabatic
condition $|\langle n_{0}|\partial_{t}
n_{\pm}\rangle|\ll|\eta_{\pm}|$ is fulfilled, the initial state
$|\psi_{1}\rangle=|n_{0}(0)\rangle$ will follow $|n_{0}(t)\rangle$
closely, and when $\theta(t)=\pi/4$ and $\beta=l\pi/2$
($l=0,\pm1,\pm2,\cdots$), we obtain the GHZ states:
$|\psi(t_{f})\rangle=|GHZ\rangle=(|\psi_{1}\rangle-e^{i\beta}|\psi_{7}\rangle)/\sqrt{2}$.
When $\beta=\pi$, it shows the most common form:
$|\psi(t_{f})\rangle=(|\psi_{1}\rangle+|\psi_{7}\rangle)/{\sqrt{2}}$.
However, this process will take quite a long time to obtain the
target state, which is undesirable.

\begin{figure}
 \scalebox{0.18}{\includegraphics {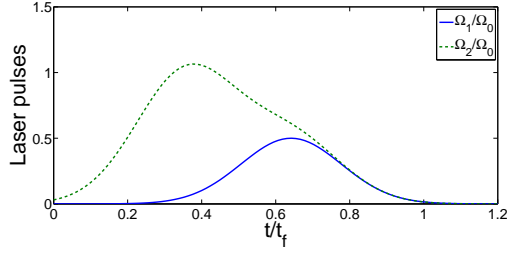}}
 \caption{Dependence on $t/t_{f}$ of $\Omega_{1}/\Omega_{0}$ and $\Omega_{3}/\Omega_{0}$.}
 \label{O1O3}
\end{figure}

\begin{figure}
 \scalebox{0.2}{\includegraphics {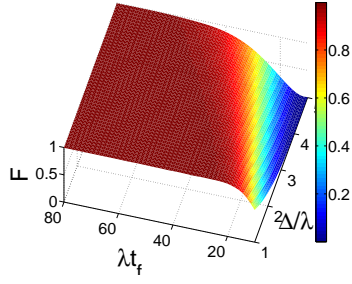}}
 \caption{The fidelity $F$ of the target state $\frac{1}{\sqrt{2}}(|\psi_{1}\rangle-|\psi_{7}\rangle)$ versus the interaction time $\lambda t_{f}$ and the
  detuning $\Delta/\lambda$.}
 \label{FDeltatf}
\end{figure}

\section{Using TQD to construct shortcuts to adiabatic passage}
The instantaneous eigenstates $|n_{k}\rangle$ ($k=0,\pm$) for the
effective Hamiltonian $H_{eff}(t)$ in section III do not satisfy the
Schr\"{o}dinger equation
$i\partial_{t}|n_{k}\rangle=H_{eff}(t)|n_{k}\rangle$. According to
Berry's general transitionless tracking algorithm \cite{MVBJpa09},
from $H_{eff}(t)$, one can reverse engineer $H(t)$ which is related
to the original Hamiltonian $H_{eff}(t)$ but drives the eigenstates
exactly. From refs.
\cite{XCARSSADCDDOJGMPrl10,MLYXLTSJSNBAPra14,XCJGMPra10} and section
II, we learn the simplest Hamiltonian $H(t)$ is derived in the form
\begin{eqnarray}\label{eq2-1}
  H(t)=i\sum_{k=0,\pm}|{\partial_{t}n_{k}(t)\rangle\langle n_{k}(t)|}.
\end{eqnarray}
Substituting eq. (\ref{eq1-5}) in eq. (\ref{eq2-1}), we obtain
\begin{eqnarray}\label{eq2-2}
  H(t)=i\dot{\theta}e^{i\beta}|\psi_{7}\rangle\langle\psi_{1}|+H.c.,
\end{eqnarray}
where
$\dot{\theta}=[\dot{\Omega}_{1}(t)\Omega_{3}(t)-\dot{\Omega}_{3}(t)\Omega_{1}(t)]/\Omega^{2}$.
Similar to ref. \cite{MLYXLTSJSNBAPra14}, for this three-atom system
in a real experiment, the Hamiltonian $H(t)$ is hard or even
impossible to be implemented in practice. We should find an APF
Hamiltonian whose effect is equivalent to $H(t)$. The model used for
the APF Hamiltonian is similar to that in Fig. \ref{model} with
three atoms trapped in a cavity, and the atomic level configuration
is shown in Fig. \ref{modelGHZ2}: the transition
$|f\rangle\leftrightarrow|e\rangle$ is non-resonantly driven by
classical field with time-dependent Rabi frequency $\tilde{\Omega}$
and detuning $\Delta$, the transition $|g_{l}\rangle\
(|g_{r}\rangle)\leftrightarrow|e\rangle$ is coupled non-resonantly
to the cavity with coupling $\lambda_{l}$ ($\lambda_{r}$) and
detuning $\Delta$. The rotating-frame Hamiltonian reads
\begin{eqnarray}\label{eq2-2a}
 H'_{I}&=&H'_{al}+H'_{ac}+H_{e},                                                  \cr\cr
 H'_{al}&=&\tilde{\Omega}_{1}(t)|e\rangle_{1}\langle f|+e^{i\beta'}\tilde{\Omega}_{3}(t)|e\rangle_{3}\langle f|+H.c.,       \cr\cr
 H'_{ac}&=&\sum_{m=1,2}{\lambda_{l}a_{l}|e\rangle_{m}\langle g_{l}|}+\sum_{n=2,3}{\lambda_{r}a_{r}|e\rangle_{n}\langle g_{r}|}+H.c., \cr\cr
 H_{e}&=&\sum_{k=1}^{3}{\Delta|e\rangle_{k}\langle e|},
\end{eqnarray}
where $\beta'$ is the phase difference between $\tilde{\Omega}_{1}$ and $\tilde{\Omega}_{3}$.
Then similar to the
approximation for the Hamiltonian from eq. (\ref{eq1-1}) to eq. (\ref{eq1-4}), we also obtain an effective
Hamiltonian for the present non-resonant system \cite{YHCYXJSQip14}
\begin{eqnarray}\label{eq2-3}
  H'_{eff}&=&[\frac{1}{\sqrt{3}}|\phi_{0}\rangle(\tilde{\Omega}_{1}(t)\langle\psi_{1}|
             +e^{i\beta'}\tilde{\Omega}_{3}(t)\langle\psi_{7}|)+H.c.]   \cr\cr
             &&+\Delta|\phi_{0}\rangle\langle\phi_{0}|.
\end{eqnarray}
By adiabatically eliminating the state $|\phi_{0}\rangle$ under the condition $\Delta\gg \tilde{\Omega}_{1}/\sqrt{3},\tilde{\Omega}_{3}/\sqrt{3}$,
we obtain the final effective Hamiltonian
\begin{eqnarray}\label{eq2-4}
  H_{fe}&=&-\frac{\tilde{\Omega}_{1}^{2}}{3\Delta}|\psi_{1}\rangle\langle\psi_{1}|
           -\frac{\tilde{\Omega}_{3}^{2}}{3\Delta}|\psi_{7}\rangle\langle\psi_{7}|               \cr\cr
           &&-\frac{e^{i\beta'}\tilde{\Omega}_{1}\tilde{\Omega}_{3}}{3\Delta}|\psi_{7}\rangle\langle\psi_{1}|
           -\frac{e^{-i\beta'}\tilde{\Omega}_{1}\tilde{\Omega}_{3}}{3\Delta}|\psi_{1}\rangle\langle\psi_{7}|.
\end{eqnarray}
Choosing $\tilde{\Omega}_{1}=\tilde{\Omega}_{3}=\tilde{\Omega}(t)$,
the first two terms of eq. (\ref{eq2-4}) can be removed, and the
Hamiltonian becomes
\begin{eqnarray}\label{eq2-5}
  \tilde{H}_{eff}=e^{i\beta'}\Omega_{x}(t)|\psi_{7}\rangle\langle\psi_{1}|+e^{-i\beta'}\Omega_{x}(t)|\psi_{1}\rangle\langle\psi_{7}|,
\end{eqnarray}
where $\Omega_{x}(t)=-\tilde{\Omega}^{2}/(3\Delta)$.
This effective Hamiltonian is equivalent to the CDD Hamiltonian $H(t)$ in eq (\ref{eq2-2}) when
\begin{eqnarray}\label{eq2-5a}
  e^{i\beta'}\Omega_{x}=ie^{i\beta}\dot{\theta}.
\end{eqnarray}
Hence, the Rabi frequencies for the APF Hamiltonian are designed
\begin{eqnarray}\label{eq2-5b}
  \tilde{\Omega}_{1}&=&\tilde{\Omega}_{3}=\sqrt{-3\Delta\dot{\theta}}, \cr
  \beta'&-&\beta=\frac{\pi}{2}+2l\pi,
\end{eqnarray}
where $l=0,\pm1,\pm2,\cdots$.

\begin{figure}
 \renewcommand\figurename{\small FIG.}
 \centering \vspace*{8pt} \setlength{\baselineskip}{10pt}
 \subfigure[]{
 \includegraphics[scale = 0.18]{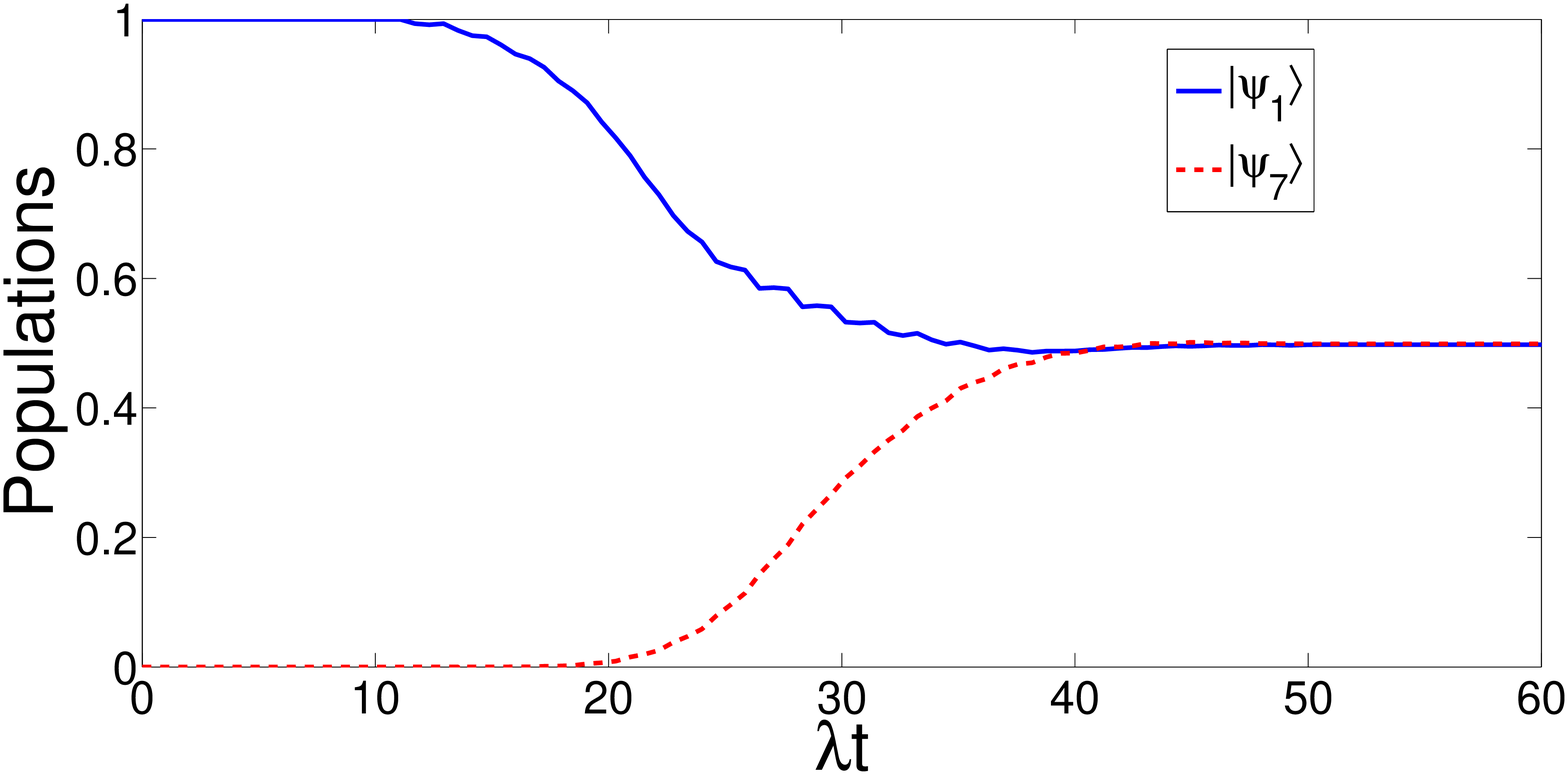}}
 \subfigure[]{
 \includegraphics[scale = 0.18]{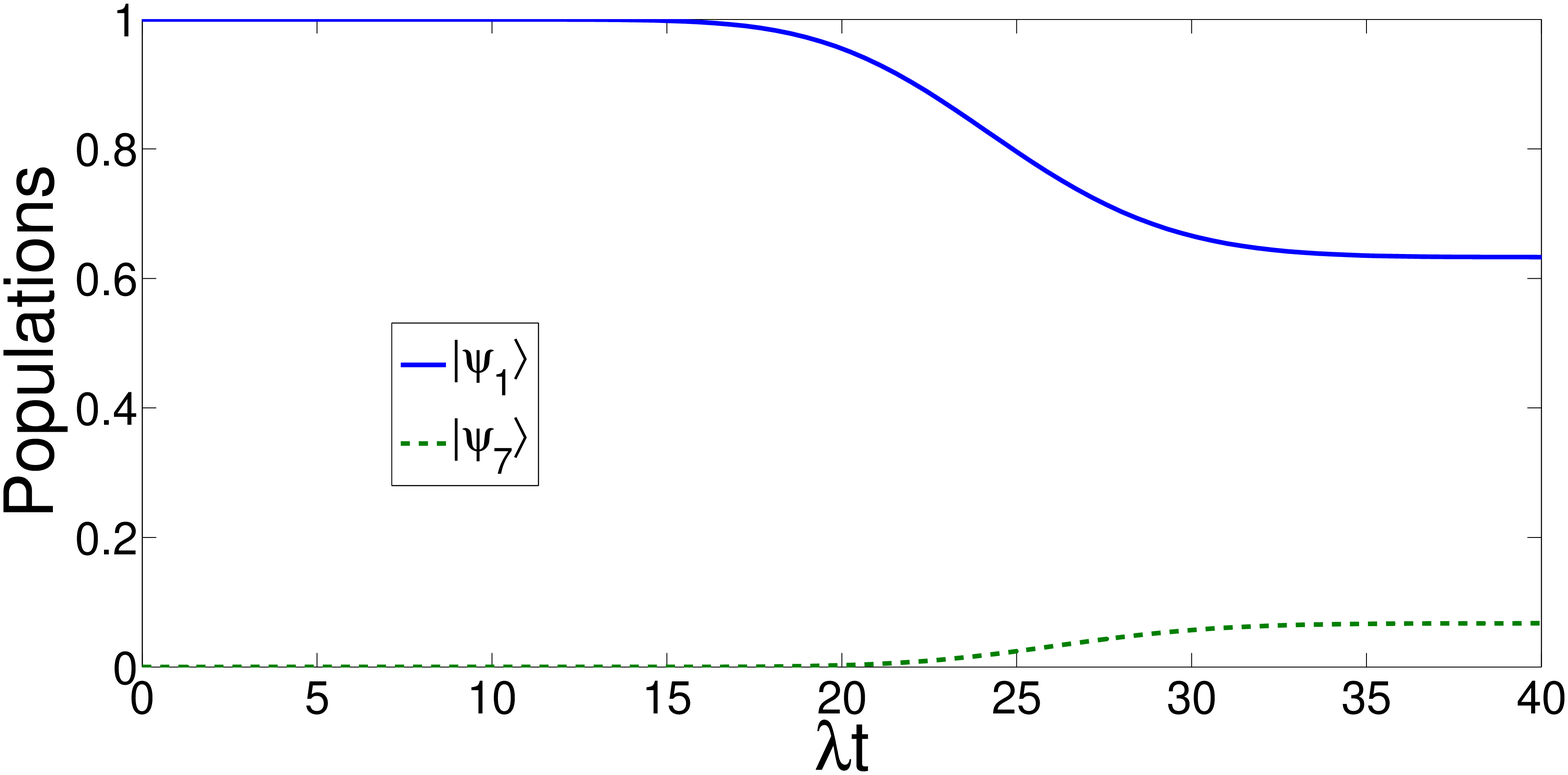}}
 \caption{
        Time evolution of the populations for the states $|\psi_{1}\rangle$ and $|\psi_{7}\rangle$ with $\Omega_{0}=0.2\lambda$, $t_{f}=35/\lambda$ and $\Delta=2.2\lambda$
        (a) governed by the APF Hamiltonian $H'_{I}(t)$,
        (b) governed by the original Hamiltonian $H_{I}(t)$.
          }
 \label{P15stapadi}
\end{figure}

\section{Fast and noise-resistant generation of the three-atom GHZ states with STAP}

\begin{figure}
 \scalebox{0.18}{\includegraphics {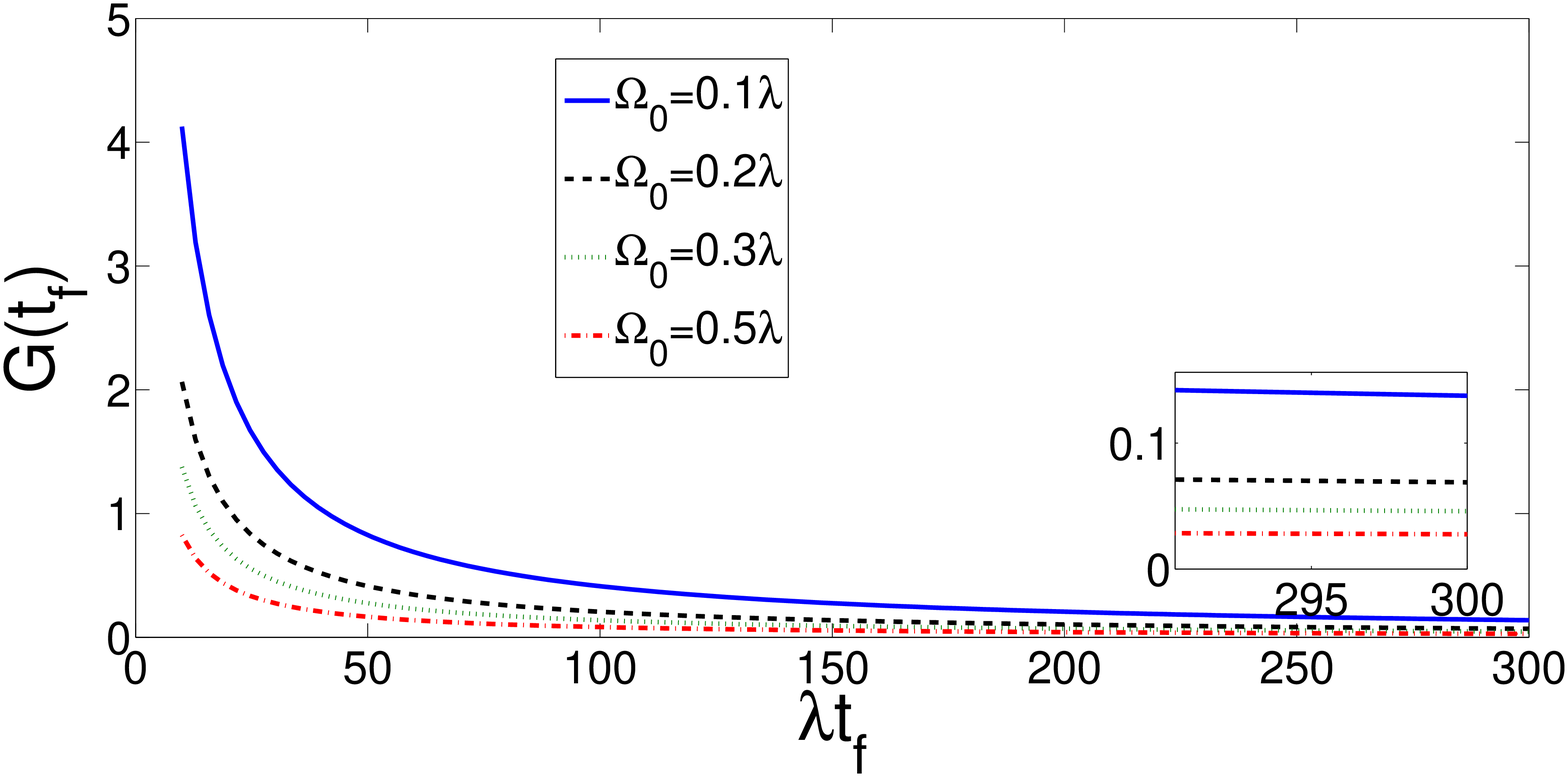}}
 \caption{The relationship between $G(t_{f})$ and $\lambda t_{f}$ for testing the adiabatic condition.}
 \label{Adicon}

 \scalebox{0.18}{\includegraphics {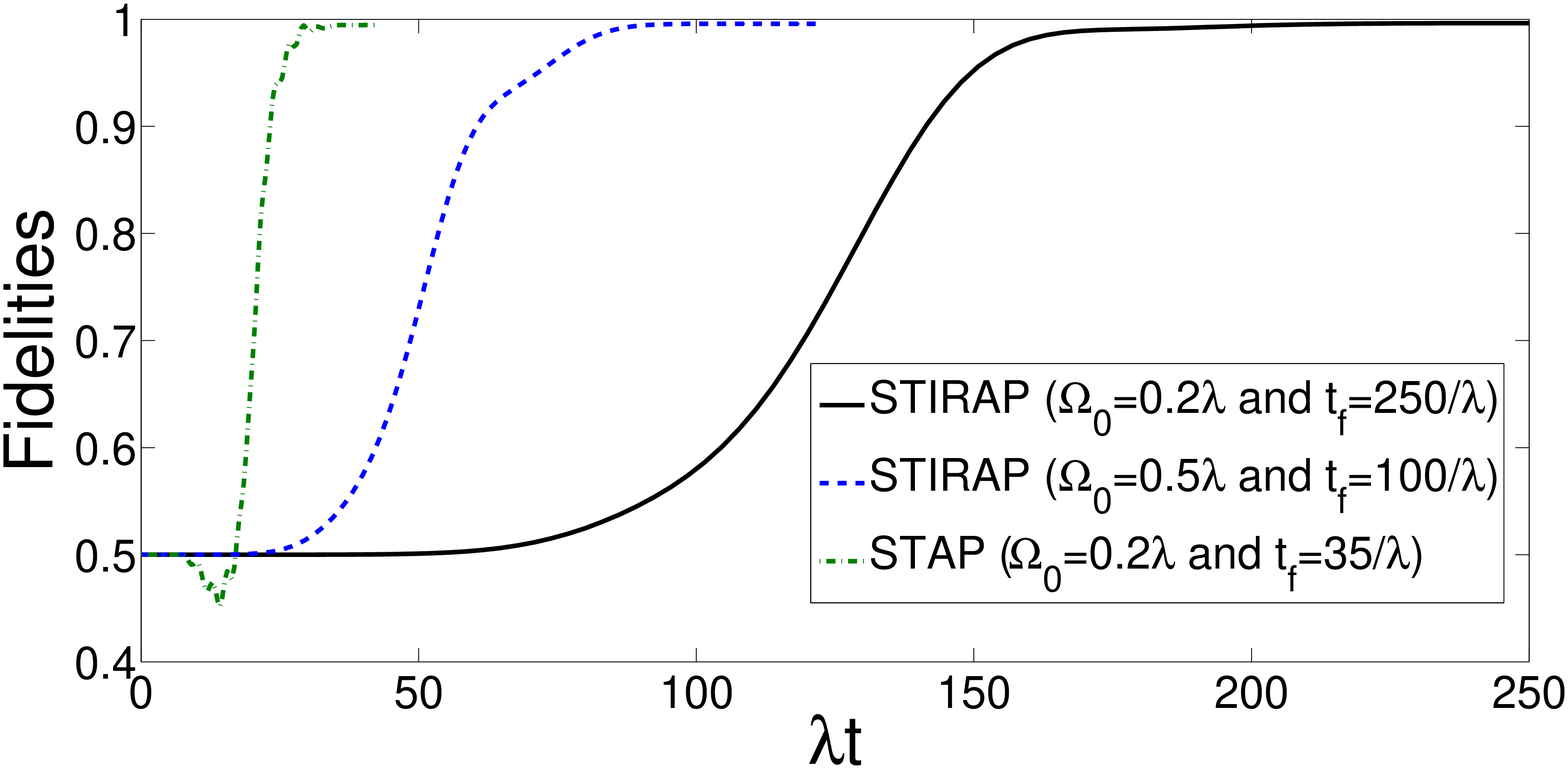}}
 \caption{The comparison between the fidelities of the three-atom GHZ state governed by the APF Hamiltonian $H'_{I}(t)$ and the original Hamiltonian $H_{I}(t)$.}
 \label{STAADI}
\end{figure}

We will show that the creation of a three-atom GHZ state governed by
$H'_{I}$ is much faster than that governed by $H_{I}$. To satisfy the
boundary condition of the fractional stimulated Raman adiabatic
passage (STIRAP),
\begin{eqnarray}\label{eq3-1}
  \lim_{t\rightarrow-\infty}\frac{\Omega_{1}(t)}{\Omega_{3}(t)}=0,\
  \lim_{t\rightarrow+\infty}\frac{\Omega_{3}(t)}{\Omega_{1}(t)}=\tan{\alpha},
\end{eqnarray}
the Rabi frequencies $\Omega_{1}(t)$ and $\Omega_{3}(t)$ in the original Hamiltonian $H_{I}(t)$ are chosen as
\begin{eqnarray}\label{eq3-2}
  \Omega_{1}(t)&=&\sin{\alpha}\Omega_{0}\exp[\frac{-(t-t_{0}-t_{f}/2)^{2}}{t_{c}^{2}}], \cr\cr
  \Omega_{3}(t)&=&\Omega_{0}\exp[\frac{-(t+t_{0}-t_{f}/2)^{2}}{t_{c}^{2}}]              \cr\cr
                  &&+\cos{\alpha}\Omega_{0}\exp[\frac{-(t-t_{0}-t_{f}/2)^{2}}{t_{c}^{2}}],
\end{eqnarray}
where $\Omega_{0}$ is the pulse amplitude, $t_{f}$ is the operation
time, and $t_{0}$, $t_{c}$ are some related parameters. In order to
create a three-atom GHZ state, the finial state
$|\psi(t_{f})\rangle$ should be
$|\psi(t_{f})\rangle=\frac{1}{\sqrt{2}}(|\psi_{1}\rangle-e^{i\beta}|\psi_{7}\rangle)$
according to eq. (\ref{eq1-5}). Therefore, we have $\tan{\alpha}=1$.
By choosing parameters for the laser pulses suitably to fulfill the
boundary condition in eq. (\ref{eq3-1}), the time-dependent
$\Omega_{1}(t)$ and $\Omega_{3}(t)$ are gotten as shown in Fig.
\ref{O1O3} with parameters $t_{0}=0.14t_{f}$ and $t_{c}=0.19t_{f}$.
For simplicity, we set $\beta=0$ in the following discussion.
Fig. \ref{FDeltatf} shows the relationship between the fidelity of
the generated three-atom GHZ state (governed by the APF Hamiltonian
$H'_{I}(t)$) and two parameters $\Delta$ and $t_{f}$ when
$\Omega_{0}=0.2\lambda$ satisfying the Zeno condition, where the
fidelity for the three-atom GHZ state is given through $F=|\langle
GHZ|\rho(t_{f})|GHZ\rangle|$ ($\rho(t_{f})$ is the density operator
of the whole system when $t=t_{f}$). We find that there is a wide
range of selectable values for parameters $\Delta$ and $t_{f}$ to
get a high fidelity of the three-atom GHZ state. The fidelity
increases with the increasing of $t_{f}$ while decreases with the
increasing of $\Delta$. It is not hard to understand, putting eq.
(\ref{eq3-2}) into eq. (\ref{eq2-5b}) and setting $t=t'\times t_{f}$,
we can find
\begin{eqnarray}\label{3-3}
  \Omega'_{0}\approx\sqrt{\frac{6\Delta}{t_{f}}},
\end{eqnarray}
where $\Omega'_{0}$ is the amplitude of $\tilde{\Omega}(t')$.
That means, in order to satisfy the Zeno condition $\tilde{\Omega}\ll
\sqrt{3}\lambda$ and the large detuning condition
$\tilde{\Omega}\ll\sqrt{3}\Delta$, the ratio $\Delta/t_{f}$ should
be small enough. Moreover, this relationship also explains the
phenomenon in Fig. \ref{FDeltatf} that to achieve a high fidelity
with a larger detuning $\Delta$, a longer interaction time $t_{f}$
is required. Then to prove the operation time required for the
creation of the three-atom GHZ state governed by $H'_{I}$ is much
shorter than that governed by $H_{I}$, we contrast the performances
of population transfer from the initial state $|\psi_{1}\rangle$
governed by the APF Hamiltonian $H'_{I}$ and that governed by the
original Hamiltonian $H_{I}$ in Fig. \ref{P15stapadi} with
$\{t_{f}=35/\lambda,\ \Omega_{0}=0.2\lambda,\ \Delta=2.2\lambda\}$.
The time-dependent population for any state $|\psi\rangle$ is given
by the relationship $P=|\langle\psi|\rho(t)|\psi\rangle|$, where
$\rho(t)$ is the corresponding time-dependent density operator. The
comparison of Figs. \ref{P15stapadi} (a) and (b) shows that with
this set of parameters, the APF Hamiltonian $H'_{I}(t)$ can govern
the evolution to achieve a near-perfect three-atom GHZ state from
state $|\psi_{1}\rangle$ in short interaction time while the
original Hamiltonian $H_{I}(t)$ can not. In fact, through solving
the adiabatic condition $|\langle
n_{0}|\partial_{t}n_{\pm}\rangle|\ll|\eta_{\pm}|$, we obtain
\begin{eqnarray}\label{eq3-4}
  |\frac{\dot{\theta}}{\sqrt{2}}|\ll |\frac{\Omega}{\sqrt{3}}|\Rightarrow \frac{f(t)}{t_{f}}\ll \Omega,
\end{eqnarray}
where $f({t})$ is a wave function whose amplitude is irrelevant to
$t_{f}$. The result shows when $\Omega_{0}$ is a constant, the
longer the operation time $t_{f}$ is, the better the adiabatic
condition is satisfied. This is proved in Fig. \ref{Adicon}. Fig.
\ref{Adicon} reveals the relationship between $G(t_{f})$ and
$\lambda t_{f}$, where
$G(t_{f})=\frac{\sqrt{3}|\dot{\theta}|}{\sqrt{2}\Omega}|_{t=0.5t_{f}}$.
From this figure, we discover that even with $\Omega_{0}=0.5\lambda$
which does not meet the Zeno condition, the operation time required
for the three-atom GHZ state generation in an adiabatic system is
longer than $100/\lambda$ (when $t_{f}=100/\lambda$,
$G(t_{f})\approx 0.08$). We also plot the fidelities of the evolved
states governed by $H'_{I}(t)$ and $H_{I}(t)$ (in different cases)
in Fig. \ref{STAADI}, with respect to the target three-atom GHZ
state. Shown in the figure, even with a large laser intensity, say,
$\Omega_{0}=0.5\lambda$, the interaction time required for creation
of the three-atom GHZ state via adiabatic passage is still much
longer than that via STAP. Generally speaking, the adiabatic
condition is satisfied much better with a relatively larger laser
intensity, while, the system would be very sensitive to the
decoherence caused by the cavity decay with a relatively large laser
intensity. This will be proved in the following.

\begin{figure}
 \renewcommand\figurename{\small FIG.}
 \centering \vspace*{8pt} \setlength{\baselineskip}{10pt}
 \subfigure[]{
 \includegraphics[scale = 0.2]{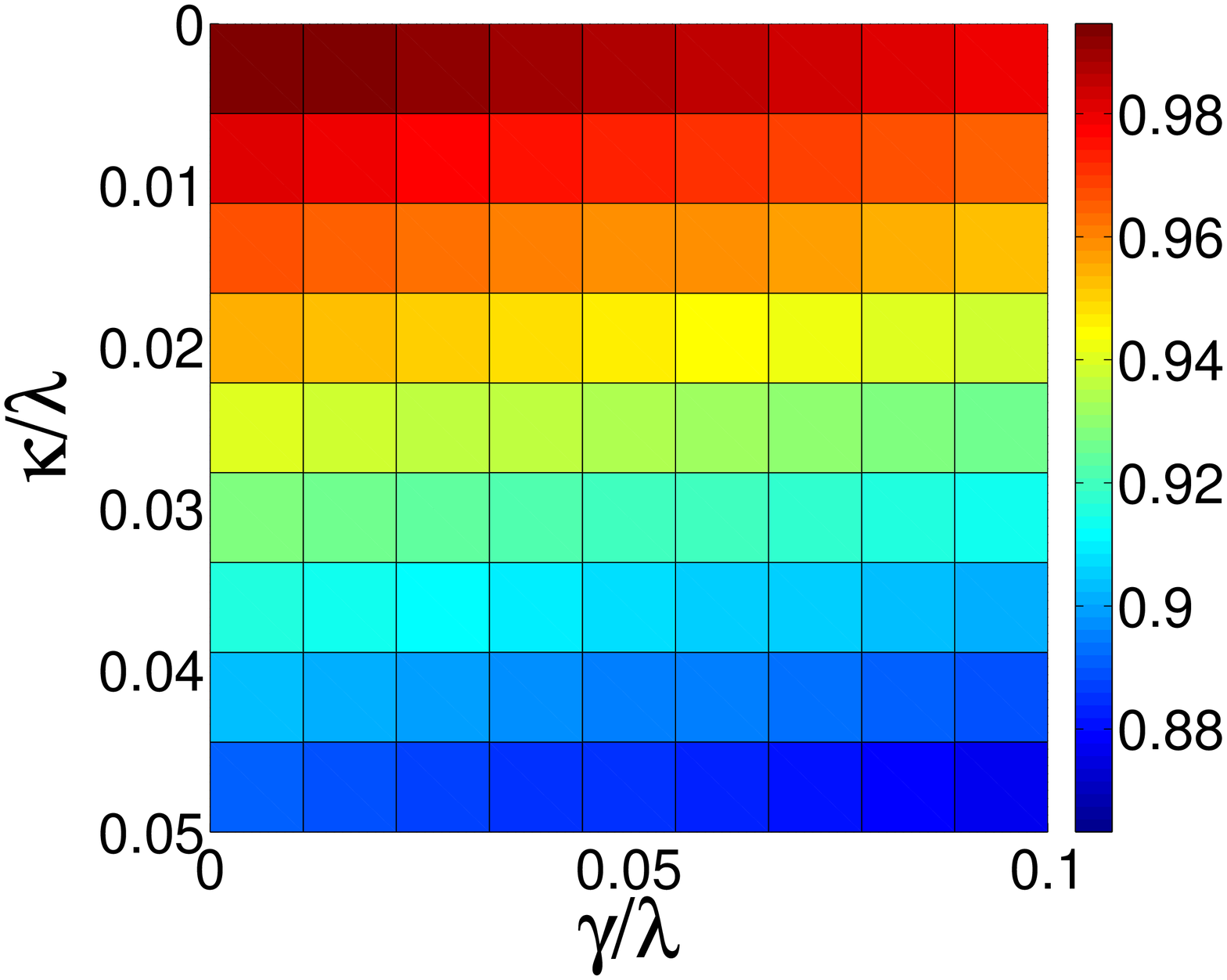}}
 \subfigure[]{
 \includegraphics[scale = 0.2]{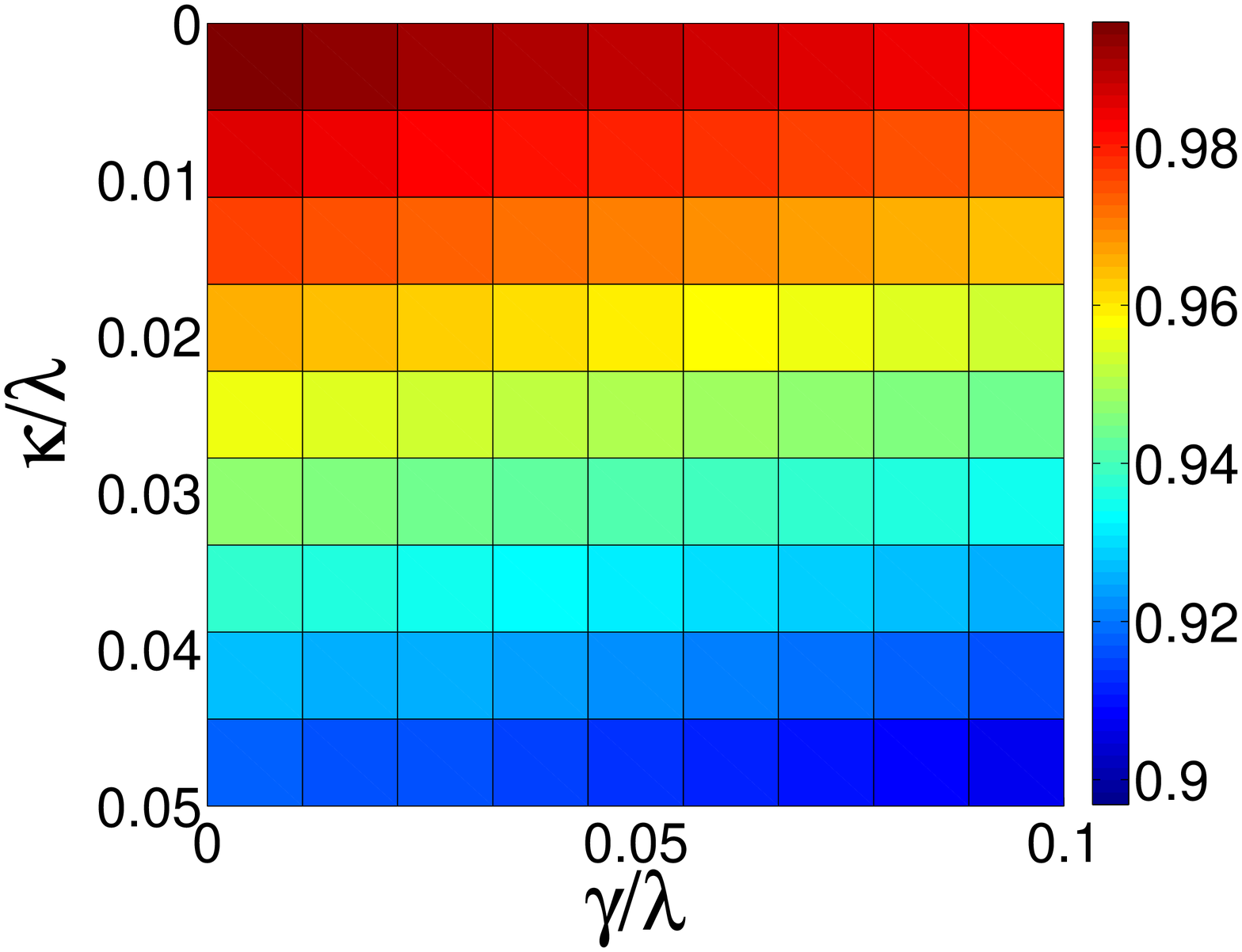}}
 \caption{
    (a) Dependences on $\kappa/\lambda$ and $\gamma/\lambda$ of the fidelity of the three-atom GHZ state governed by the APF Hamiltonian $H'_{I}(t)$
        when $t_{f}=35/\lambda$, $\Delta=2.2\lambda$ and $\Omega_{0}=0.2\lambda$.
    (b) Dependences on $\kappa/\lambda$ and $\gamma/\lambda$ of the fidelity of the three-atom GHZ state governed by the original Hamiltonian
        $H_{I}(t)$ when $t_{f}=100/\lambda$ and $\Omega_{0}=0.5\lambda$.
          }
 \label{Fkr}
\end{figure}

Once the dissipation is considered, the evolution of the system can be modeled by a master
equation in Lindblad form,
\begin{eqnarray}\label{eq3-5}
  \dot{\rho}=i[\rho,H]+\sum_{k}[L_{k}\rho L_{k}^{\dag}-\frac{1}{2}(L_{k}^{\dag}L_{k}\rho+\rho L_{k}^{\dag}L_{k})],
\end{eqnarray}
where $L_{k}$'s are the Lindblad operators. For both the resonant and non-resonant systems, there are
eight Lindblad operators governing the dissipation:
\begin{eqnarray}\label{eq3-6}
  L_{1}^{\kappa}&=&\sqrt{\kappa_{l}}a_{l},                    \ \ \ \ \ \  \
  L_{2}^{\kappa}=\sqrt{\kappa_{r}}a_{r},                          \cr
  L_{3}^{\gamma}&=&\sqrt{\gamma_{1}}|f\rangle_{1}\langle e|,      \
  L_{4}^{\gamma}=\sqrt{\gamma_{2}}|g_{l}\rangle_{1}\langle e|,   \cr
  L_{5}^{\gamma}&=&\sqrt{\gamma_{3}}|g_{l}\rangle_{2}\langle e|,
  L_{6}^{\gamma}=\sqrt{\gamma_{4}}|g_{r}\rangle_{2}\langle e|,   \cr
  L_{7}^{\gamma}&=&\sqrt{\gamma_{5}}|f\rangle_{3}\langle e|,    \
  L_{8}^{\gamma}=\sqrt{\gamma_{6}}|g_{r}\rangle_{3}\langle e|,
\end{eqnarray}
where $\kappa_{l}$ and $\kappa_{r}$ are the decays of the cavity
modes, and $\gamma_{n}$ ($n=1,2,\cdots,6$) are the spontaneous
emissions of atoms. For simplicity, we assume
$\kappa_{l}=\kappa_{r}=\kappa$, and $\gamma_{n}=\gamma/2$. Fig.
\ref{Fkr} (a) shows the fidelity of the three-atom GHZ state
governed by the APF Hamiltonian $H'_{I}$ versus these two noise
resources with $\Omega_{0}=0.2\lambda$, $\Delta=2.2\lambda$, and
$t_{f}=35/\lambda$. It turns out that the present shortcut scheme
with this set of parameters is much more sensitive to the cavity
decays than the spontaneous emissions. Ref. \cite{YHCYXQQCJSPra14}
contributes to understanding this phenomenon, in fact, with this set
of parameters, the Zeno condition for the non-resonant system is not
ideally fulfilled because shortening the time implies an energy cost
\cite{XCJGMPra10,XCETJGMPra10} (in this system, the energy cost denotes
requiring relative-large laser intensities). Known from ref.
\cite{YHCYXQQCJSPra14}, destroying the Zeno condition
slightly is also helpful to achieve the target state in a much shorter
interaction time. However, if the Zeno condition has not been satisfied
very well, the intermediate states including the cavity-excited
states would be populated during the evolution, which causes that
the system is sensitive to the cavity decays. However, we can find
in Fig. \ref{Fkr} (b) which shows fidelity of the three-atom GHZ
state governed by original Hamiltonian $H_{I}$ with
$\Omega_{0}=0.5\lambda$ and $t_{f}=100/\lambda$ in the presence of
decoherence, with large laser intensities, the adiabatic scheme is
also sensitive to the cavity decays as we mentioned above.
The comparison of these two figures drops a result that the present
shortcut scheme is almost the same with the adiabatic one in
restraining the decoherence.

\begin{figure}
 \renewcommand\figurename{\small FIG.}
 \centering \vspace*{8pt} \setlength{\baselineskip}{10pt}
 \subfigure[]{
 \includegraphics[scale = 0.2]{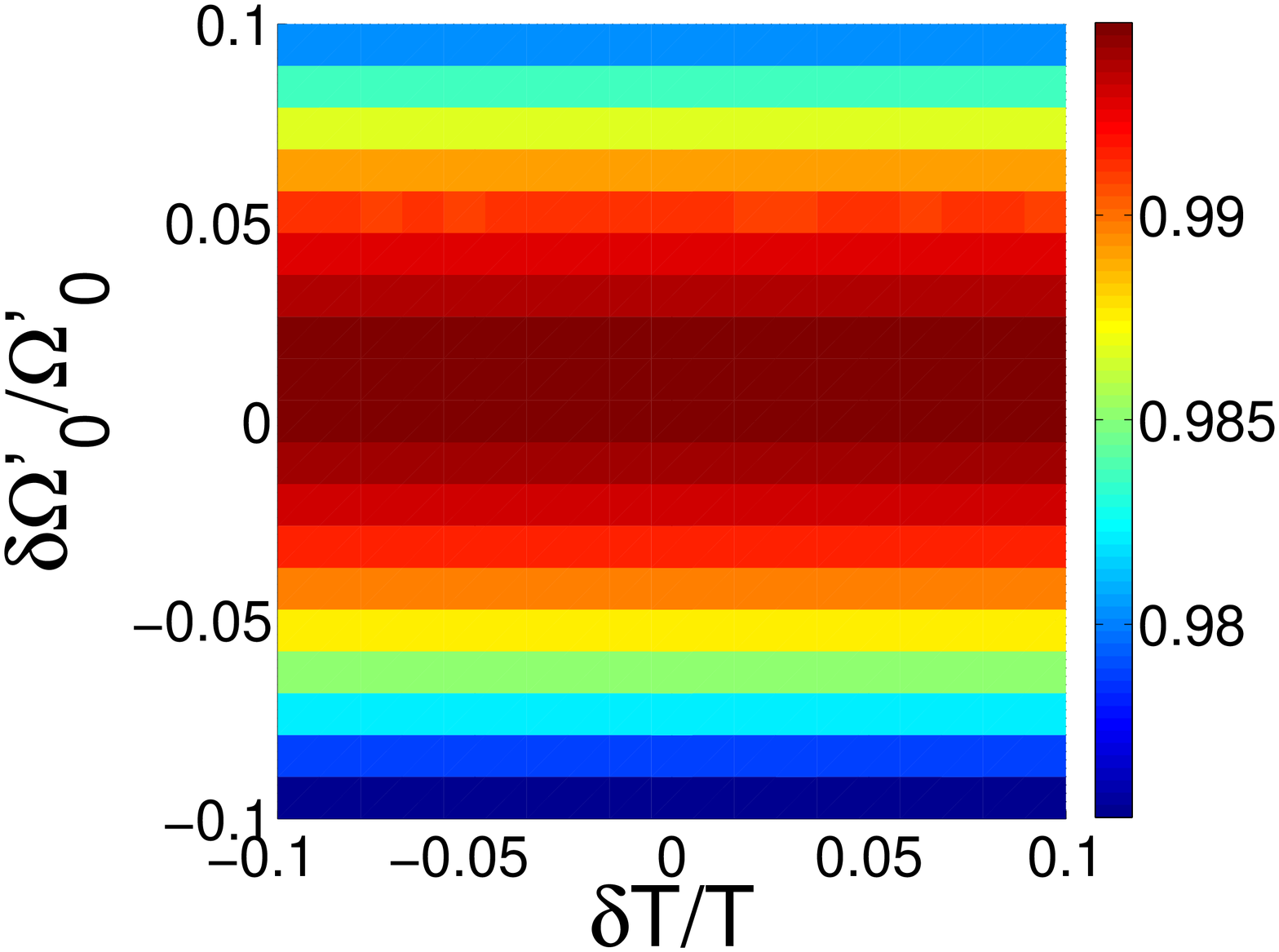}}
 \subfigure[]{
 \includegraphics[scale = 0.2]{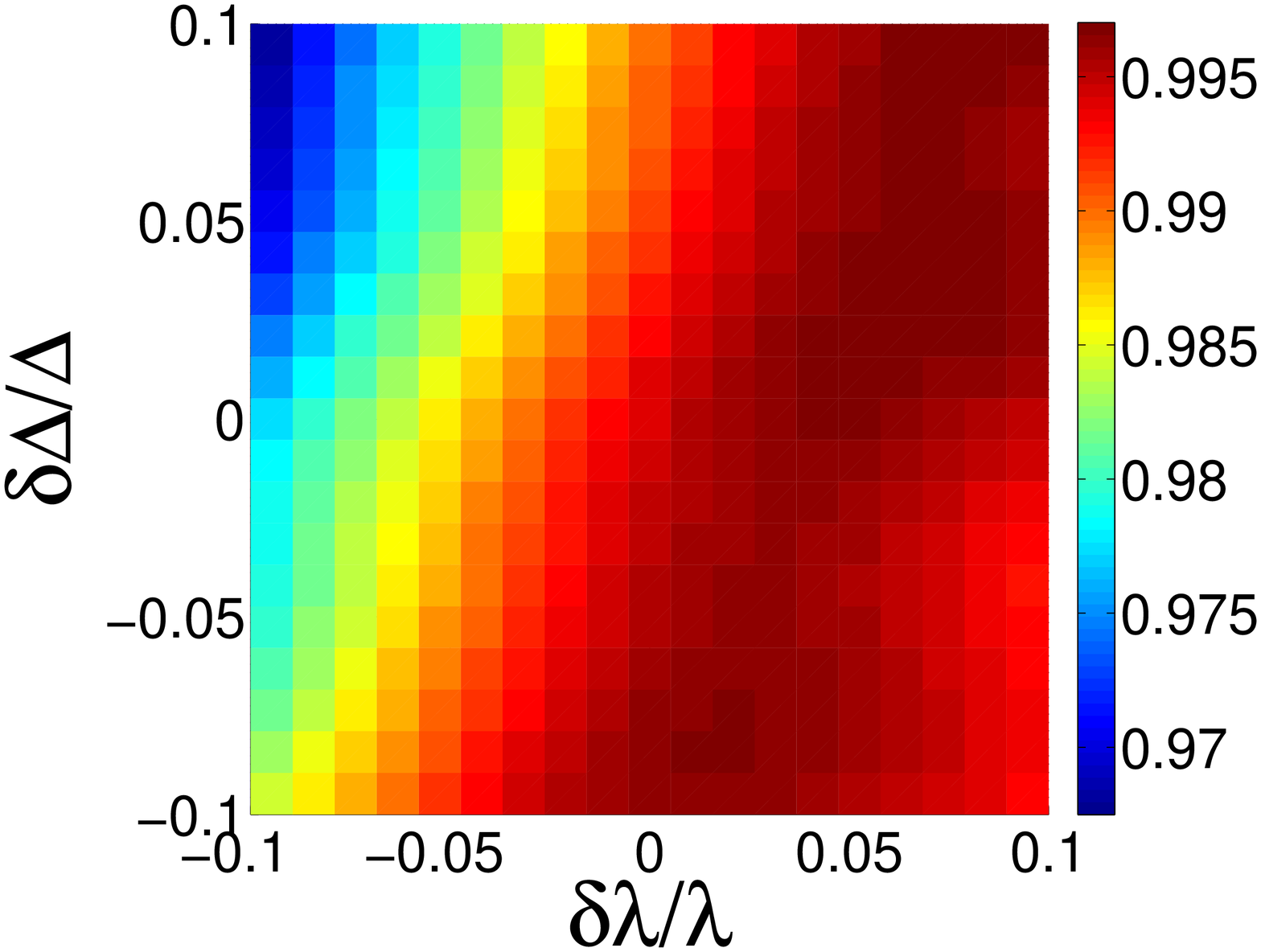}}
 \caption{
    The fidelity $F$ of the GHZ state versus the variations of (a) $T$ and
     $\Omega'_{0}$. (b) $\lambda$ and $\Delta$.
          }
 \label{Fdelta}
\end{figure}

The robustness against operational imperfection is also a main
factor for the feasibility of the scheme because most of the
parameters are hard to accurately achieve in experiment. Therefore,
we define $\delta x=x'-x$ as the deviation of any parameter $x$,
where $x'$ is the actual value and $x$ is the ideal value. Then in
Fig. \ref{Fdelta} (a) we plot the fidelity of the GHZ state versus
the variations in total operation time $T$ ($T=1.2t_{f}$) and laser
amplitude $\Omega'_{0}$, and in Fig \ref{Fdelta} (b) we plot the
fidelity of the GHZ state versus the variations in coupling
$\lambda$ and detuning $\Delta$. As shown in the figures, the scheme
is robust against all of these variations. Any deviation $\delta
x/x=10\%$ ($x\in\{T,\Omega'_{0},\lambda,\Delta\}$) causes a
reduction less than $3\%$ in the fidelity.

In a real experiment, the cesium atoms which have been cooled and
trapped in a small optical cavity in the strong-coupling regime
\cite{JMJRBADBAKHCNDMSKHJKPrl03,JYDWVHJKPrl99} can be
used in this scheme. With a set of cavity QED parameters
$\lambda=750\times 2\pi$ MHz, $\kappa=3.5\times 2\pi$ MHz, and
$\gamma=2.62\times 2\pi$ MHz
\cite{SMSTJKKJVKWGEWHJKPra05,MJHFGSLBMBPNat06}, the fidelity of the
three-atom GHZ state in this paper is 98.24\%. Thus, the scheme is
robust and might be promising within the limits of current
technology.

\section{Conclusion}
We have presented a promising method to construct shortcuts to
adiabatic passage (STAP) for a three-atom system to generate GHZ
states in the cavity QED system. Through using quantum Zeno dynamics
and ``transitionless quantum driving'', we are free to simplify a
complicated Hamiltonian and choose the laser pulses to construct
shortcuts in multi-qubit system to implement the fast quantum
information processing. Numerical simulation demonstrates that the
scheme is fast and robust against the decoherence caused by both
atomic spontaneous emission, photon leakage and operational
imperfection. The deficiency is that the present scheme might be
sensitive to the cavity decays because of some inevitable factors.
Compared with the previous shortcut methods, this method obviously
works better at entanglement generation in multi-qubit systems. In
fact, any quantum system whose Hamiltonian is possible to be
simplified into the form in eq. (\ref{eq1-4}), the shortcut can be
constructed with the same method presented in this paper. For
example, similar to refs. \cite{WALLFWOe12,SYHYXJSNBAJosab13} for
the generation of the multiparticle GHZ states in an
atom-fiber-cavity combined system, we can shorten the operation time
using the same method in the following steps: (1) With the help of
the quantum Zeno dynamics, we can simplify the Hamiltonian of the
single-excitation subspace into an effective Hamiltonian
$H_{eff}(t)$ with the form in eq. (\ref{eq1-4}). (2) For this
effective Hamiltonian, by using TQD, we construct the CDD
Hamiltonian $H(t)$ that speeds up the adiabatic process. (3) Similar
to section IV, we find out the corresponding non-resonant system
(the APF Hamiltonian) whose effective Hamiltonian
$\tilde{H}_{eff}(t)$ has the form in eq. (\ref{eq2-5}). (4) Making
$\tilde{H}_{eff}(t)=H(t)$, the parameters for the APF Hamiltonian
are determined and the shortcut is constructed. Then the APF
Hamiltonian would govern the system to achieve the same final result
as the adiabatic process governed by the original Hamiltonian with a
much shorter operation time. This might lead to a useful step toward
realizing fast and noise-resistant quantum information processing
for multi-qubit systems in current technology.

\section*{ACKNOWLEDGEMENT}

  This work was supported by the National Natural Science Foundation of China under Grants No.
11105030 and No. 11374054, the Foundation of Ministry of Education
of China under Grant No. 212085, and the Major State Basic Research
Development Program of China under Grant No. 2012CB921601.


\begin{thebibliography}{999}
  \bibitem{XCILARDGOJGMPra10} Chen,~X., Lizuain,~I., Ruschhaupt,~A., Gu\'{e}ry-Odelin,~D., \& Muga,~J.~G. 
       Shortcut to adiabatic passage in two- and three-level atoms. Phys. Rev. Lett.
      \textbf{105}, 123003 (2010).
  \bibitem{ETSISMGMMACDGOARXCJGMAmop13} Torrontegui,~E., Ib\'{a}\~{n}ez,~S., Mart\'{i}nez-Garaot,~S.,
                                        Modugno,~M., del~Campo,~A., Gu\'{e}-Odelin,~D., Ruschhaupt,~A., Chen,~X., \& Muga,~J.~G.
      Chapter 2 - Shortcuts to Adiabaticity. Adv. Atom. Mol. Opt. Phys.
      \textbf{62}, 117-169 (2013).
  \bibitem{AdCPrl13} del~Campo,~A.
      Shortcuts to adiabaticity by counterdiabatic driving. Phys. Rev. Lett.
      \textbf{111}, 100502 (2013).
  \bibitem{SMKNPrspca10Pra11} Masuda,~S. \& Nakamura,~K.
      Acceleration of adiabatic quantum dynamics in electromagnetic fields. Phys. Rev. A
      \textbf{84}, 043434 (2011).
  \bibitem{YHCYXQQCJSPra14} Chen,~Y.~H., Xia,~Y., Chen,~Q.~Q., \& Song,~J.
      Efficient shortcuts to adiabatic passage for fast population transfer in multiparticle systems. Phys. Rev. A
      \textbf{89}, 033856 (2014).
  \bibitem{MLYXLTSJSNBAPra14} Lu,~M., Xia,~Y., Shen,~L.~T., Song,~J., \& An,~N.~B.
      Shortcuts to adiabatic passage for population transfer and maximum entanglement creation between two atoms in a cavity. Phys. Rev. A
      \textbf{89}, 012326 (2014).
  \bibitem{MLYXLTSJSLp14} Lu,~M., Xia,~Y., Shen,~L.~T., \& Song,~J.,
      An effective shortcut to adiabatic passage for fast quantum state transfer in a cavity quantum electronic dynamics system. Laser Phys.
      \textbf{24}, 105201 (2014).
  \bibitem{YHCYXQQCJSLpl14} Chen,~Y.~H., Xia,~Y., Chen,~Q.~Q., \& Song,~J.
      Shortcuts to adiabatic passage for multiparticles in distant cavities: applications to fast and noise-resistant quantum population transfer,
      entangled states' preparation and transition. Laser Phys. Lett.
      \textbf{11}, 115201 (2014).
  \bibitem{YHYXQQCJSarxiv14} Chen,~Y.~H., Xia,~Y., Chen,~Q.~Q., \& Song,~J.
      Fast and noise-resistant implementation of quantum phase gates and creation of quantum entangled states. Phys. Rev. A
      \textbf{91}, 012325(2015).
  \bibitem{JGMXCARDGOJpb09} Muga,~J.~G., Chen,~X., Ruschhaup,~A., \& Gu\'{e}ry-Odelin,~D.
      Frictionless dynamics of Bose每Einstein condensates under fast trap variations. J. Phys. B
      \textbf{42}, 241001 (2009).
  \bibitem{XCARSSADCDDOJGMPrl10} Chen,~X., Ruschhaupt,~A., Schmidt,~S., del~Campo,~A., Gu\'{e}ry-Odelin,~D., \& Muga,~J.~G.
      Fast optimal frictionless atom cooling in harmonic traps: Shortcut to adiabaticity. Phys. Rev. Lett.
      \textbf{104}, 063002 (2010).
  \bibitem{XCJGMPra10} Chen,~X. \& Muga,~J.~G.
      Transient energy excitation in shortcuts to adiabaticity for the time-dependent harmonic oscillator. Phys. Rev. A
      \textbf{82}, 053403 (2010).
  \bibitem{JFSPCGLPVNjp11} Schaff,~J.~F., Capuzzi,~P., Labeyrie,~G., \& Vignolo,~P.
      Shortcuts to adiabaticity for trapped ultracold gases. New J. Phys.
      \textbf{13}, 113017 (2011).
  \bibitem{ETSIXCARDGOJGMPra11} Torrontegui,~E., Ib\'{a}\~{n}ez,~S., Chen,~X., Ruschhaupt,~A., Gu\'{e}ry-Odelin,~D., \& Muga,~J.~G.
      Fast atomic transport without vibrational heating. Phys. Rev. A
      \textbf{83}, 013415 (2011).
  \bibitem{XCETDSJSLJGMPra11} Chen,~X., Torrontegui,~E., Stefanatos,~D., Li,~J.~S., \& Muga,~J.~G.
      Optimal trajectories for efficient atomic transport without final excitation. Phys. Rev. A
      \textbf{84}, 043415 (2011).
  \bibitem{ETXCMMSSARJGMNjp12} Torrontegui,~E., Chen,~X., Modugno,~M., Schmidt,~S., Ruschhaupt,~A., \& Muga,~J.~G.
      Fast transport of Bose每Einstein condensates. New J. Phys.
      \textbf{14}, 013031 (2012).
  \bibitem{YLLAWZDWPra11} Li,~Y., Wu,~L.~A., \& Wang,~Z.~D.
      Fast ground-state cooling of mechanical resonators with time-dependent optical cavities. Phys. Rev. A
      \textbf{83}, 043804 (2011).
  \bibitem{AdCPra11} del~Campo,~A.
      Frictionless quantum quenches in ultracold gases: A quantum-dynamical microscope. Phys. Rev. A
      \textbf{84}, 031606(R) (2011);
      Fast frictionless dynamics as a toolbox for low-dimensional Bose-Einstein condensates. Eur. Phys. Lett.
      \textbf{96}, 60005 (2011).
  \bibitem{ARXCDAJGMNjp12} Ruschhaupt,~A., Chen,~X., Alonso,~D., \& Muga,~J.~G.
      Optimally robust shortcuts to population inversion in two-level quantum systems. New J. Phys.
      \textbf{14}, 093040 (2012).
  \bibitem{JFSXLSPVGLPra10} Schaff,~J.~F., Song,~X.~L., Vignolo,~P., \& Labeyrie,~G.
      Fast optimal transition between two equilibrium states. Phys. Rev. A
      \textbf{82}, 033430 (2010).
  \bibitem{JFSXLSPCPVGLEpl11} Schaff,~J.~F., Song,~X.~L., Capuzzi,~P., Vignolo,~P., \& Labeyrie,~G.
      Shortcut to adiabaticity for an interacting Bose-Einstein condensate. Eur. Phys. Lett.
      \textbf{93}, 23001 (2011).
  \bibitem{AWFZTRSTDOKTMHKSFSKUPPrl12} Walther,~A., Ziesel,~F., Ruster,~T., Dawkins,~S.~T., Ott,~K., Hettrich,~M., Singer,~K., Schmidt-Kaler,~F., \& Poschinger,~U.
      Number-Theoretic Nature of Communication in Quantum Spin Systems. Phys. Rev. Lett.
      \textbf{109}, 050502 (2012).
  \bibitem{SYTXCOl12} Tseng,~S.~Y. \& Chen,~X.
      Engineering of fast mode conversion in multimode waveguides. Opt. Lett.
      \textbf{37}, 5118-5120 (2012).
  \bibitem{XCETJGMPra10} Chen,~X., Torrontegui,~E., \& Muga,~J.~G.
      Lewis-Riesenfeld invariants and transitionless quantum driving. Phys. Rev. A
      \textbf{83}, 062116 (2011).

  \bibitem{HRLWBRJmp69} Lewis,~H.~R. \& Riesenfeld,~W.~B. 
      An exact quantum theory of the time-dependent harmonic oscillator and of a charged particle in a time-dependent electromagnetic field. J. Math. Phys.
      \textbf{10}, 1458-1473 (1969).
  \bibitem{MVBJpa09} Berry,~M.~V.
      Transitionless quantum driving. J. Phys. A
      \textbf{42}, 365303 (2009).
  \bibitem{MGBMVNMPHEADCRFVGRMONatPhys12} Bason,~M.~G., Viteau,~M., Malossi,~N., Huillery,~P., Arimondo,~E.,
                                          Ciampini,~D., Fazio,~R., Giovannetti,~V., Mannella,~R., \& Morsch,~O. 
      High-fidelity quantum driving. Nat. Phys.
      \textbf{8}, 147-152 (2012).
  \bibitem{MDSARJpca03} Demirplak,~M. \& Rice,~S.~A. 
      Adiabatic population transfer with control fields. J. Phys. Chem. A
      \textbf{107}, 9937-9945 (2003).
  \bibitem{MDSARJcp08} Demirplak,~M. \& Rice,~S.~A. 
      On the consistency, extremal, and global properties of counterdiabatic fields. J. Chem. Phys.
      \textbf{129}, 154111 (2008).
  \bibitem{SIXCETJGMARPrl12} Ib\'{a}\~{n}ez,~S., Chen,~X., Torrontegui,~E., Muga,~J.~G., \& Ruschhaupt,~A. 
      Multiple schr\"{o}dinger pictures and dynamics in shortcuts to adiabaticity. Phys. Rev. Lett.
      \textbf{109}, 100403 (2012).
  \bibitem{SMGETXCJGMPra14} Mart\'{i}nez-Garaot,~S., Torrontegui,~E., Chen,~X., \& Muga,~J.~G. 
      Shortcuts to adiabaticity in three-level systems using Lie transforms. Phys. Rev. A
      \textbf{89}, 053408 (2014).
  \bibitem{ETSMGJGMPra14} Torrontegui,~E., Mart\'{i}nez-Garaot,~S. \& Muga,~J.~G. 
      Hamiltonian engineering via invariants and dynamical algebra. Phys. Rev. A
      \textbf{89}, 043408 (2014).
  \bibitem{TOKMMjp14} Opatrn\'{y},~T. \& M{\o}lmer,~K. 
      Partial suppression of nonadiabatic transitions. New J. Phys.
      \textbf{16}, 015025 (2014).
  \bibitem{DFJJJCJP07} James,~D.~F. \& Jerke,~J. 
      Effective Hamiltonian theory and its applications in quantum information. Can. J. Phys.
      \textbf{85}, 625-632 (2007).
  \bibitem{PKHWTHAZMAKPrl95} Kwiat,~P., Weinfurter,~H., Herzog,~T., Zeilinger,~A., \& Kasevich,~M.~A. 
      Interaction-free measurement. Phys. Rev. Lett.
      \textbf{74}, 4763-4766 (1995).
  \bibitem{PFSPPrl02}Facchi,~ P. \& Pascazio,~S. 
      Quantum Zeno subspaces. Phys. Rev. Lett.
      \textbf{89}, 080401 (2002).
  \bibitem{DMGMHASAZAjp90} Greenberger,~D.~M., Horne,~M.~A., Shimony,~A., \& Zeilinger,~A. 
      Bell＊s theorem without inequalities. Am. J. Phys.
      \textbf{58}, 1131-1143 (1990).
  \bibitem{SBZPrl01} Zheng,~S.~B. 
      One-Step synthesis of multiatom Greenberger-Horne-Zeilinger states. Phys. Rev. Lett.
      \textbf{87}, 230404 (2001).

  \bibitem{YHCYXJSQip14} Chen,~Y.~H., Xia,~Y. \& Song,~J. 
      Deterministic generation of singlet states for N-atoms in coupled cavities via quantum Zeno dynamics. Quantum info. Proc.
      \textbf{13}, 1857-1877 (2014).
  \bibitem{JYDWVHJKPrl99} Ye,~J.,Vernooy,~ D.~W., \& Kimble,~H.~J.
      Trapping of Single Atoms in Cavity QED. Phys. Rev. Lett.
      \textbf{83}, 4987-4990 (1999).
  \bibitem{JMJRBADBAKHCNDMSKHJKPrl03} McKeever,~J., Buck,~J.~R., Boozer,~A.~D., Kuzmich,~A., N\"{a}gerl,~H.~C., Stamper-Kurn,~D.~M., \& Kimble,~H.~J.
      State-Insensitive cooling and trapping of single atoms in an optical cavity. Phys. Rev. Lett.
      \textbf{90}, 133602 (2003).
  \bibitem{SMSTJKKJVKWGEWHJKPra05} Spillane,~S.~M., Kippenberg,~T.~J., Vahala,~K.~J., Goh,~K.~W., Wilcut,~E., \& Kimble,~H.~J.
      Ultrahigh-$Q$ toroidal microresonators for cavity quantum electrodynamics. Phys. Rev. A
      \textbf{71}, 013817 (2005).
  \bibitem{MJHFGSLBMBPNat06} Hartmann,~M.~J., Brand\~{a}o,~F.~G.~S.~L., \& Plenio,~M.~B. 
      Strongly interacting polaritons in coupled arrays of cavities. Nat. Phys.
     \textbf{2}, 849-855 (2006).
  \bibitem{WALLFWOe12} Li,~W.~A. \& Wei,~L.~F. 
      Controllable entanglement preparations between atoms in spatially-separated cavities via quantum Zeno dynamics. Opt. Express
      \textbf{20}, 13440-13450 (2012).
  \bibitem{SYHYXJSNBAJosab13} Hao,~S.~Y., Xia,~Y., Song,~J, \& An,~N.~B.
      One-step generation of multiatom Greenberger每Horne每Zeilinger states in separate cavities via adiabatic passage. J. Opt. Soc. Am. B
      \textbf{30}, 468-474 (2013).

\end{thebibliography}
\end{document}